\documentstyle[preprint,aps,floats,epsf]{revtex}

\begin{document}

\tightenlines 

\catcode`@=11
\def\references{%
\ifpreprintsty
\bigskip\bigskip
\hbox to\hsize{\hss\large \refname\hss}%
\else
\vskip24pt
\hrule width\hsize\relax
\vskip 1.6cm
\fi
\list{\@biblabel{\arabic{enumiv}}}%
{\labelwidth\WidestRefLabelThusFar  \labelsep4pt %
\leftmargin\labelwidth %
\advance\leftmargin\labelsep %
\ifdim\baselinestretch pt>1 pt %
\parsep  4pt\relax %
\else %
\parsep  0pt\relax %
\fi
\itemsep\parsep %
\usecounter{enumiv}%
\let\p@enumiv\@empty
\def\theenumiv{\arabic{enumiv}}%
}%
\let\newblock\relax %
\sloppy\clubpenalty4000\widowpenalty4000
\sfcode`\.=1000\relax
\ifpreprintsty\else\small\fi
}
\catcode`@=12


\preprint{\font\fortssbx=cmssbx10 scaled \magstep2
\hbox to \hsize{
\hbox{\fortssbx University of Wisconsin - Madison}
\hfill$\vcenter{\hbox{\bf MADPH-97-1020}
                \hbox{\bf hep-ph/9711328}
                \hbox{November 1997}}$ }
}
 
\title{\vspace{.5in}
Detecting the Higgs Bosons of Minimal Supergravity \\ with Muon Pairs}
 
\author{V. Barger and Chung Kao}

\address{
Department of Physics, University of Wisconsin, 
Madison, WI 53706, USA}

\maketitle

\bigskip

\begin{abstract}

The prospects at the CERN LHC are investigated 
for the discovery via decays into muon pairs of neutral Higgs bosons 
in the minimal supergravity model.  
Promising results are found for the CP-odd pseudoscalar ($A^0$) 
and the heavier CP-even scalar ($H^0$) Higgs bosons 
at large $\tan\beta \equiv v_2/v_1$. 
For $\tan\beta \agt 10$ and 100 GeV $\alt m_A,m_H \alt$ 550 GeV, 
the $A^0$ and the $H^0$ masses could be precisely reconstructed 
from the dimuon invariant mass.

\end{abstract}

%


\newpage

\section{Introduction}

In the Standard Model (SM) of electroweak interactions, 
the masses of gauge bosons and fermions are generated 
by a scalar field doublet.
After spontaneous symmetry breaking, 
a neutral CP-even Higgs boson ($h^0_{SM}$) remains as a physical particle.
The scalar sector of the SM is unstable to radiative corrections 
that arise when the SM is coupled to ultraheavy degrees of freedom 
such as those in a grand unified theory (GUT).
Therefore, there is great interest in extensions of the SM that can solve 
this problem. 

A supersymmetry (SUSY) between fermions and bosons provides 
a natural explanation of the Higgs mechanism 
for electroweak symmetry breaking (EWSB) 
in the framework of a grand unified theory (GUT) 
and preserves the elementary nature of Higgs bosons. 
For the particle content 
of the minimal supersymmetric standard model (MSSM) \cite{MSSM}, 
the evolution of gauge couplings 
by renormalization group equations (RGEs) \cite{RGE} 
is consistent with a grand unified scale 
at $M_{\rm GUT} \sim 2\times 10^{16}$ GeV 
and an effective SUSY mass scale in the range 
$M_Z < M_{\rm SUSY} \alt 10$ TeV \cite{Unification}. 
With a large top quark Yukawa coupling ($Y_t$) to a Higgs boson 
at the GUT scale, radiative corrections drive 
the corresponding Higgs boson mass squared parameter negative, 
spontaneously breaking the electroweak symmetry 
and naturally explaining the origin of the electroweak scale.
In the minimal supersymmetric GUT with a large $Y_t$, 
there is an infrared fixed point (IRFP) \cite{BBO,IRFP} 
at low $\tan\beta$; the top quark mass is correspondingly predicted 
to be $m_t = (200 \; {\rm GeV}) \sin\beta$ \cite{BBO}, 
and thus $\tan\beta \simeq 1.8$ for $m_t = 175$ GeV. 
At high $\tan\beta$, another quasi-IRFP solution ($dY_t/dt \simeq 0$) exists 
at $\tan\beta \sim 56$.

The Higgs sector of a supersymmetric theory must contain 
at least two $SU(2)$ doublets \cite{Guide} for anomaly cancellation. 
In the minimal supersymmetric standard model (MSSM), 
the Higgs sector has two doublets $\phi_1$ and $\phi_2$ 
that couple to the $t_3 = -1/2$ and $t_3 = +1/2$ fermions, respectively.  
After spontaneous symmetry breaking, there remain five physical Higgs bosons:
a pair of singly charged Higgs bosons $H^{\pm}$,
two neutral CP-even scalars $H^0$ (heavier) and $h^0$ (lighter),
and a neutral CP-odd pseudoscalar $A^0$.
The Higgs potential is constrained by supersymmetry 
such that all tree-level Higgs boson masses and couplings 
are determined by just two independent parameters,  
commonly chosen to be the mass of the CP-odd pseudoscalar ($m_A$) 
and the ratio of vacuum expectation values (VEVs) of Higgs fields 
($\tan\beta \equiv v_2/v_1$). 

Extensive studies have been made of the prospects for the detection
of MSSM Higgs bosons at the LHC 
\cite{HGG,Neutral,KZ,Z2Z2,CMS,ATLAS,ATLAS2}.
Most studies have focused on the SM decay modes 
$\phi \to \gamma\gamma$ ($\phi = H^0, h^0$ or $A^0$) 
and $\phi \to ZZ$ or $ZZ^*\to 4l$ ($\phi = H^0$ or $h^0$) 
that are detectable above background.
For $\tan\beta$ close to one, 
the detection modes $A^0 \to Zh^0 \to l^+l^- b\bar{b}$ 
or $l^+l^- \tau\bar{\tau}$ \cite{AZh} 
and $H^0 \to h^0 h^0 \to \gamma\gamma b\bar{b}$ \cite{ATLAS2} 
may provide promising channels to simultaneously discover 
two Higgs bosons of the MSSM. 
There are regions of parameter space where rates for Higgs boson
decays to SUSY particles are large and dominant.
While these decays reduce the rates for the standard modes, 
making conventional detection of Higgs bosons more difficult, 
they also open up a number of new promising modes 
for Higgs detection \cite{Z2Z2}.

For large $\tan\beta$, 
the $\tau\bar{\tau}$ decay mode \cite{KZ,CMS,ATLAS} 
is a promising discovery channel for the $A^0$ and the $H^0$ in the MSSM.
It was also suggested that neutral Higgs bosons might be observable via
their $b\bar{b}$ decays \cite{Dandi,DGV} in a large region
of the ($m_A,\tan\beta$) plane,
provided that sufficient $b$-tagging capability can be achieved.
However, simulations with ATLAS detector performance concluded that 
detection of the $b\bar{b}$ channel would be difficult \cite{hbb}.

Recently, the muon pair decay mode was proposed \cite{Nikita,CMS} 
and confirmed \cite{ATLAS2} to be a promising discovery channel for
the $A^0$ and the $H^0$.
Although the $\mu\bar{\mu}$ channel has a small branching fraction, 
this is compensated by the much better achievable mass resolution 
with muon pairs.
For large $\tan\beta$, the muon pair discovery mode might 
be the only channel at the LHC that allows precise reconstruction
of the $A^0$ and the $H^0$ masses. 

In supergravity (SUGRA) models \cite{SUGRA}, 
supersymmetry is broken in a hidden sector 
with SUSY breaking communicated to the observable sector 
through gravitational interactions, 
leading naturally but not necessarily \cite{Non-universal} 
to a common scalar mass ($m_0$), a common gaugino mass ($m_{1/2}$), 
a common trilinear coupling ($A_0$) and a common bilinear coupling ($B_0$) 
at the GUT scale.
Through minimization of the Higgs potential, 
the $B$ coupling parameter of the superpotential  
and the magnitude of the Higgs mixing parameter $\mu$ 
are related to the ratio of VEVs of Higgs fields ($\tan\beta \equiv v_2/v_1$) 
and to the mass of the $Z$ boson ($M_Z$).
The SUSY particle masses and couplings at the weak scale 
can be predicted by the evolution of RGEs 
from the unification scale \cite{BBO,SUGRA2}. 
%

For $\tan\beta$ close to 1.8, the lower IRFP, 
masses of the CP-odd pseudoscalar ($m_A$) 
and the heavier CP-even scalar ($m_H$) are large 
over most of the minimal supergravity parameter space. 
Then the couplings of the lighter scalar $h^0$ 
are similar to those of the SM Higgs boson; 
the $h^0$ could be the only neutral Higgs boson 
observable at the CERN LHC. 
For large $\tan\beta$, 
the $A^0$ and the $H^0$ can become light \cite{Baer,Madison} 
and potentially visible at the LHC.

Recent measurements of the $b \to s\gamma$ decay rate 
by the CLEO \cite{CLEO} and LEP collaborations \cite{LEP} 
place constraints on the parameter space 
of the minimal supergravity model \cite{bsg}.
It was found that $b \to s\gamma$ excludes 
most of the minimal supergravity (mSUGRA) parameter space 
when $\tan\beta$ is large and $\mu > 0$ \cite{bsg}. 
Therefore, we will choose $\mu < 0$ in our analysis.
However, our results are almost independent of the sign of $\mu$.

The remainder of this article addresses 
the prospects of discovering the neutral Higgs bosons 
in the mSUGRA model via their decays into muon pairs at the LHC. 
The cross section and branching fractions are presented in Section II.
The observability of the dimuon signals is discussed in Section III 
for conservative assumption about the detector mass resolution. 
Promising conclusions are drawn in Section IV.

\section{Cross Section and Branching Fraction}

We evaluate SUSY mass spectra and couplings 
in the minimal supergravity model 
with four parameters: $m_0$, $m_{1/2}$, $A_0$ and $\tan\beta$, 
and with the sign of the Higgs mixing parameter $\mu$. 
Since $A_0$ mainly affects the masses of third generation sfermions, 
which do not significantly affect our analysis, 
we take $A_0 = 0$ in our calculations. 

The mass matrix of the charginos in the weak eigenstates 
($\tilde{W}^\pm$, $\tilde{H}^\pm$) has the following form \cite{BBO}
\begin{equation}
M_C=\left( \begin{array}{c@{\quad}c}
M_2 & \sqrt{2}M_W\sin \beta \\
\sqrt{2}M_W\cos \beta & -\mu
\end{array} \right)\;.
\label{eq:xino}
\end{equation}
This mass matrix is not symmetric and must be diagonalized 
by two matrices \cite{MSSM}.
The sign of the $\mu$ contribution in Eq.~(\ref{eq:xino}) 
establishes our sign convention for $\mu$. 

We calculate the masses and couplings in the Higgs sector 
with one-loop corrections from both the top and the bottom Yukawa interactions 
in the RGE-improved one-loop effective potential \cite{Higgs} 
at the scale $Q = \sqrt{m_{\tilde{t}_L}m_{\tilde{t}_R}}$. 
With this scale choice, the numerical value 
of the CP-odd Higgs boson mass ($m_A$) 
at large $\tan\beta$ \cite{Baer,Madison} 
is relatively insensitive to the exact scale choice 
and the loop corrections to $m_A$ are small 
compared to the tree level contribution.
In addition, when this high scale is used, 
the RGE improved one-loop corrections approximately 
reproduce the dominant two-loop perturbative calculation 
of the mass of the lighter CP-even Higgs scalar ($m_h$).  
Our numerical values of $m_h$ are very close to the results of 
Ref.~\cite{Two-Loop} where somewhat different scales higher than $M_Z$ 
have been adopted in evaluating the effective potential.
 
The cross section of 
$pp \to \phi \to \mu\bar{\mu} +X$ ($\phi = A^0, H^0$, or $h^0$) 
is calculated for the two dominant subprocesses 
$gg \to \phi$ and $gg \to \phi b\bar{b}$, 
and multiplied by the branching fraction of the Higgs decay into muon pairs
$B(\phi \to \mu\bar{\mu})$.
The parton distribution functions of CTEQ3L \cite{CTEQ}
are used to evaluate the $pp \to \phi +X$ cross section 
with $\Lambda_4 = 0.177$ GeV and 
$Q = M_{gg}$ = the invariant mass of the gluons. 
We take $M_Z = 91.19$ GeV, $\sin^2\theta_W = 0.2319$, 
$M_W = M_Z \cos\theta_W$, 
$m_b(pole) = 4.8$ GeV, and $m_t(pole) = 175$ GeV.


At the LHC energy, 
the SM Higgs boson is produced dominantly from gluon fusion; 
vector boson fusion is also relevant if the Higgs boson is heavy.
In the MSSM, gluon fusion ($gg \to \phi$) is the major source
of neutral Higgs bosons for $\tan\beta$ less than about 4.
If $\tan\beta$ is larger than about 10,
neutral Higgs bosons in the MSSM are dominantly produced
from $b$-quark fusion ($b\bar{b} \to \phi$) \cite{Duane}.
The cross section of $gg \to \phi b\bar{b}$
is a good approximation to the `exact' cross section \cite{Duane}
of $b\bar{b} \to \phi$ for $M_\phi$ less than about 500 GeV.
Since the Yukawa couplings of $\phi b\bar{b}$ are enhanced by $1/\cos\beta$,
the production rate of neutral Higgs bosons is usually
enhanced at large $\tan\beta$.
For $m_A$ larger than about 150 GeV, the couplings of the lighter scalar $h^0$
to gauge bosons and fermions become similar to those of the SM Higgs boson. 
Then gluon fusion is the major source of the $h^0$ 
even if $\tan\beta$ is large.


The QCD radiative corrections to the subprocess $gg \to \phi$ 
are substantial \cite{QCD1,QCD2}; 
the corrections to $gg \to \phi b\bar{b}$ are still to be evaluated.
To be conservative, 
we take a K-factor of 1.5 for the contribution from $gg \to \phi$ 
and a K-factor of 1.0 for the contribution from $gg \to \phi b\bar{b}$. 
For the dominant Drell-Yan background \cite{Nikita,CMS,ATLAS2}, 
we adopt the well known K-factor from reference \cite{Collider}.


With QCD radiative corrections to $\phi \to b\bar{b}$ \cite{Manuel}, 
the branching fraction of $\phi \to \mu\bar{\mu}$ is about 
$m_\mu^2/3 m_b^2(m_b) \sim 2\times 10^{-4}$ 
when the $b\bar{b}$ mode dominates Higgs decays, 
where 3 is a color factor of the quarks 
and $m_b(m_b)$ is the running mass at the scale $m_b$.  
The branching fraction of $h^0 \to \mu\bar{\mu}$ 
is always about $2 \times 10^{-4}$.
The branching fractions for $A^0 \to \mu\bar{\mu}$ and $H^0 \to \mu\bar{\mu}$ 
are always in the range $1.5 - 2.5 \times 10^{-4}$ when $\tan\beta \agt 10$, 
even when $A^0$ and $H^0$ can decay into SUSY particles.
For $m_A$ less than about 80 GeV, the $H^0$ decays dominantly into
$h^0 h^0$, $A^0 A^0$ and $Z A^0$.

In Figures 1(a) and 1(b), 
we present the cross section of the MSSM Higgs bosons 
at the LHC, $pp \to \phi \to \mu\bar{\mu} +X$, as a function of $m_A$ 
for $m_{\tilde{q}} = m_{\tilde{g}} = -\mu = 1$ TeV, 
(a) $\tan\beta = 15$ and (b) $\tan\beta = 40$.
As $\tan\beta$ increases, 
the production cross section is enhanced because for $\tan\beta \agt 10$, 
the production cross section is dominated by $gg \to \phi b\bar{b}$
and enhanced by the $\phi b\bar{b}$ Yukawa coupling.
Also shown is the same cross section for 
the SM Higgs boson $h^0_{SM}$ with $m_{h_{SM}} = m_A$.
For $m_{h_{SM}} > 140$ GeV, the SM $h^0_{SM}$ mainly decays into gauge bosons, 
and the branching fraction $B(h^0_{SM} \to \mu\bar{\mu})$ 
drops sharply in this mass region. 

We present the cross section of $pp \to \phi \to \mu\bar{\mu} +X$
as a function of $\tan\beta$ for $\mu < 0$, $m_0 = 300 \;\; {\rm GeV}$ 
and various values of $m_{1/2}$ in Figure 2.
The regions in Fig. 2 with dark shading denote the parts 
of the parameter space that do not have the lightest neutralino ($\chi^0_1$) 
as the lightest supersymmetric particle (LSP).
As $\tan\beta$ increases, the production cross section is enhanced because 
(i) $m_A$ is reduced; and 
(ii) for $\tan\beta \agt 10$, the production cross section is enhanced 
by the $\phi b\bar{b}$ Yukawa coupling. 
Also shown in Fig. 2 are the pseudoscalar mass ($m_A$) versus $\tan\beta$ 
and the invariant mass distribution ($d\sigma/dM_{\mu\bar{\mu}}$) 
of the background from the Drell-Yan process at $M_{\mu\bar{\mu}} = m_A$. 

In Figure 3, 
we show the cross section of $pp \to \phi \to \mu\bar{\mu} +X$ in fb, 
as a function of $m_{1/2}$, with $\sqrt{s} = 14$ TeV,
for $\mu > 0$, $m_0 = 300$ GeV, and
(a) $\tan\beta = 1.8$, (b) $\tan\beta = 10$,
(c) $\tan\beta = 35$, and (d) $\tan\beta = 50$. 
Also shown are $m_A$ versus $m_{1/2}$ and 
the invariant mass distribution ($d\sigma/dM_{\mu\bar{\mu}}$)
of the background from the Drell-Yan process at $M_{\mu\bar{\mu}} = m_A$. 
The regions in Fig. 3 with dark shading denote the parts 
of the parameter space that do not satisfy 
the following theoretical requirements: 
tachyon free, and the lightest neutralino as the LSP.
For $\tan\beta \sim 1.8$, $m_A$ is very large 
and the cross section of the $A^0,H^0$ signal 
is much smaller than the background.
For $\tan\beta \agt 35$, the cross section of the $A^0,H^0$ signal is greatly 
enhanced and can become slightly larger than the background.
For $\tan\beta \sim 50$ and $m_{1/2} < 200$ GeV, 
$m_A$ can become close to $m_Z$ 
and a peak appears in the invariant mass distribution 
of the background from the Drell-Yan process.

\section{Observability at the LHC}

We define the signal to be observable 
if the $N\sigma$ lower limit on the signal plus background is larger than 
the corresponding upper limit on the background \cite{HGG,Brown}, namely,
\begin{eqnarray}
L (\sigma_s+\sigma_b) - N\sqrt{ L(\sigma_s+\sigma_b) } > 
L \sigma_b +N \sqrt{ L\sigma_b }
\end{eqnarray}
which corresponds to
\begin{eqnarray}
\sigma_s > \frac{N^2}{L} \left[ 1+2\sqrt{L\sigma_b}/N \right]
\end{eqnarray}
Here $L$ is the integrated luminosity, 
$\sigma_s$ is the cross section of Higgs signal, 
and $\sigma_b$ is the background cross
section within a bin of width $\pm\Delta M_{\mu\bar{\mu}}$ centered
at $M_\phi$; $N = 2.32$ corresponds to a $99\%$ confidence level
and $N = 2.5$  corresponds to a 5$\sigma$ signal.
We take the integrated luminosity $L$ to be 300 fb$^{-1}$ \cite{ATLAS2}. 

To study the observability of the muon discovery mode,
we consider the background from the Drell-Yan (DY) process,
$q\bar{q} \to Z,\gamma \to \mu\bar{\mu}$, which is the dominant background. 
We take $\Delta M_{\mu\bar{\mu}}$ to be the larger of
the ATLAS muon mass resolution (about $2\%$ of the Higgs bosons mass) 
\cite{ATLAS,ATLAS2} or the Higgs boson width. 
The CMS mass resolution will be better than $2\%$ of $m_\phi$ for 
$m_\phi \alt$ 500 GeV \cite{CMS,Nikita}. 
Therefore, the observability will be better for the CMS detector. 
The minimal cuts applied are (1) $p_T(\mu) > 20$ GeV and 
(2) $|\eta(\mu)| < 2.5$ for both the signal and background.
For $m_A \agt$ 130 GeV, $m_A$ and $m_H$ are almost degenerate 
while for $m_A \alt$ 100 GeV $m_A$ and $m_h$ are very close 
to each other \cite{Nikita,CMS}.
Therefore, we add up the cross sections 
of the $A^0$ and the $h^0$ for $m_A \le 100$ GeV 
and those of the $A^0$ and the $H^0$ for $m_A > 100$ GeV.

The 5$\sigma$ discovery contours for the MSSM Higgs bosons 
at $\sqrt{s} =$ 14 TeV 
with an integrated luminosity $L = 300 \;\; {\rm fb}^{-1}$ 
are shown in Figs. 1(c) and 1(d) for 
$m_{\tilde{q}} = m_{\tilde{g}} = -\mu = 1$ TeV
and $m_{\tilde{q}} = m_{\tilde{g}} = -\mu = 300$ GeV. 
The discovery region of $H^0 \to \mu\bar{\mu}$ 
is slightly enlarged for a smaller $m_A$, 
but the observable region of $h^0 \to \mu\bar{\mu}$ is slightly reduced 
because the lighter top squarks make the $H^0$ and the $h^0$ lighter; 
also the $H^0 b\bar{b}$ coupling is enhanced 
while the $h^0 b\bar{b}$ coupling is reduced.

In Figure 4, we present the LHC discovery contours 
in the minimal supergravity model, for 
(a) the $m_{1/2}$ versus $\tan\beta$ plane with $m_0 = 150$ GeV,
(b) the $m_{1/2}$ versus $\tan\beta$ plane with $m_0 = 500$ GeV,
(c) the $m_{1/2}$ versus $m_0$ plane with $\tan\beta = 15$, and
(d) the $m_{1/2}$ versus $m_0$ plane with $\tan\beta = 40$.
The discovery region is the part of the parameter space 
between the curve of square symbol and the dash line. 
The QCD radiative corrections to background from the Drell-Yan process
are included.
The regions in Fig. 4 with dark shading denote the parts 
of the parameter space that do not satisfy 
the following theoretical requirements: 
electroweak symmetry breaking (EWSB), tachyon free, 
and the lightest neutralino as the LSP.
Also shown are the mass contours for $m_A = 100$ GeV (dashed), 
500 GeV (solid) and 1000 GeV (dot-dashed).
For $\tan\beta \agt 40$, it should be possible to observe 
the $A^0$ or the $H^0$ with a mass up to about 550 GeV.

There are a couple of interesting points to note: 
(i) an increase in $\tan\beta$ leads to a larger $m_h$ 
but a reduction in $m_A$ and $m_H$; 
(ii) an increase in $m_0$ or in $m_{1/2}$ raises $m_A$, $m_H$ 
and masses of the other scalars significantly. 

\section{Conclusions}
 
The muon pair decay mode is a very promising channel
to discover the neutral Higgs bosons of minimal supersymmetry 
and minimal supergravity.
The discovery region of the $\mu\bar{\mu}$ mode 
is slightly smaller than the $\tau\bar{\tau}$ channel 
but $\mu\bar{\mu}$ allows precise reconstruction
of the Higgs boson masses.
The $A^0$ and $H^0$ should be observable in a large region of parameter space
with $\tan\beta \agt 10$.
The $h^0$ should be observable in a region with $m_A < 120$ GeV
and $\tan\beta \agt 5$.

The observable regions of the parameter space are found to be
\begin{eqnarray}
m_0 = 150 \;\; {\rm GeV}: & & \;\;\;\, m_{1/2} \alt 400 \;\;{\rm GeV}
\;\; {\rm and} \;\; \tan\beta \agt 12 \nonumber \\
m_0 = 500 \;\; {\rm GeV}: & & \;\;\;\, m_{1/2} \alt 1 \;\;{\rm TeV}
\;\; {\rm and} \;\; \tan\beta \agt 28
\end{eqnarray}
For two specific choices of large $\tan\beta$, the observable regions are 
\begin{eqnarray}
\tan\beta = 15: & & \;\;\;\, m_{1/2} \alt 200 \;\;{\rm GeV}
\;\; {\rm and} \;\; m_0 \alt 200 \;\; {\rm GeV} \nonumber \\
\tan\beta = 40: & & \;\;\;\, m_{1/2} \alt 600 \;\;{\rm GeV}
\;\; {\rm and} \;\; m_0 \alt 800 \;\; {\rm GeV} 
\end{eqnarray}
All observable regions are nearly independent of the sign of $\mu$.

For $m_A \agt 200$ GeV and $\tan\beta > 25$, $L =$ 10 fb$^{-1}$
would be sufficient \cite{Nikita} to obtain Higgs boson signals
with a statistical significance larger than 7.
For $M_{\mu\bar{\mu}}$ close to the $M_Z$, the signal is marginal 
because it appears on the shoulder of the huge $Z$ peak. 
Adequate subtraction procedures are required to extract the signal 
in this region.

\section*{Acknowledgments}

We are grateful to Nikita Stepanov for beneficial discussions. 
This research was supported in part by the U.S. Department of Energy 
under Grant No. DE-FG02-95ER40896 
and in part by the University of Wisconsin Research Committee 
with funds granted by the Wisconsin Alumni Research Foundation.
 
%


\begin{figure}
\centering\leavevmode
\epsfxsize=6in\epsffile{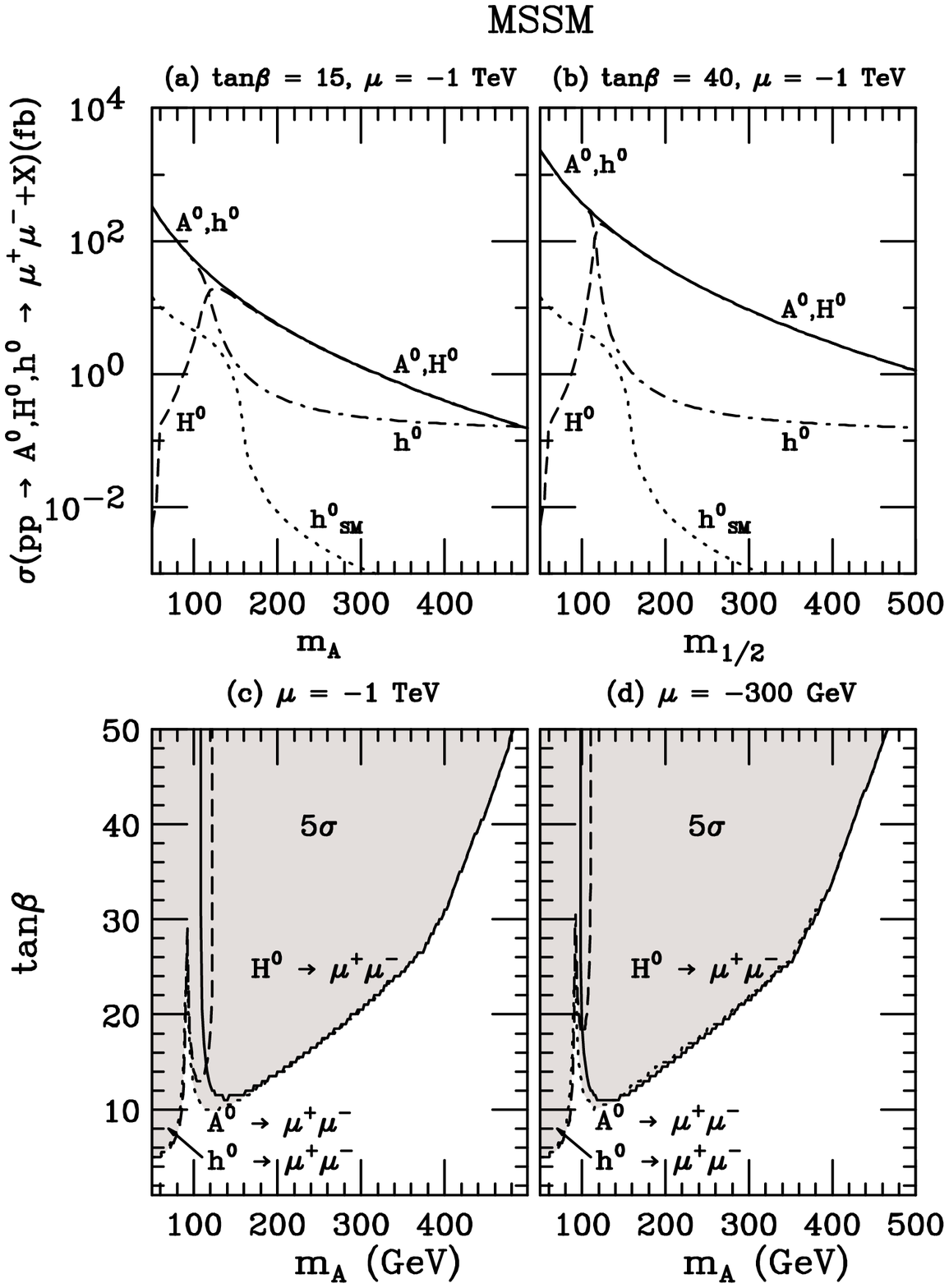}

\caption[]{
The total cross section of $pp \to A^0,H^0,h^0 \to \mu\bar{\mu} +X$ in fb 
at $\sqrt{s} = 14$ TeV, as a function of $m_A$ in the MSSM, 
for $m_{\tilde{g}} = m_{\tilde{q}} = -\mu = 1$ TeV, 
(a) $\tan\beta = 15$ and (b) $\tan\beta = 40$.
Also shown are the cross section for the SM Higgs boson 
with $m_{h_{SM}} = m_A$ (dotted). 
The 5$\sigma$ contours at the LHC with $L =$ 300 fb$^{-1}$ 
are shown for 
(c) $m_{\tilde{g}} = m_{\tilde{q}} = -\mu = 1$ TeV, and 
(d) $m_{\tilde{g}} = m_{\tilde{q}} = -\mu = 300$ GeV. 
The discovery region is the part of the parameter space with light shading.
\label{fig:MSSM}
}\end{figure}
%

\begin{figure}
\centering\leavevmode
\epsfxsize=6in\epsffile{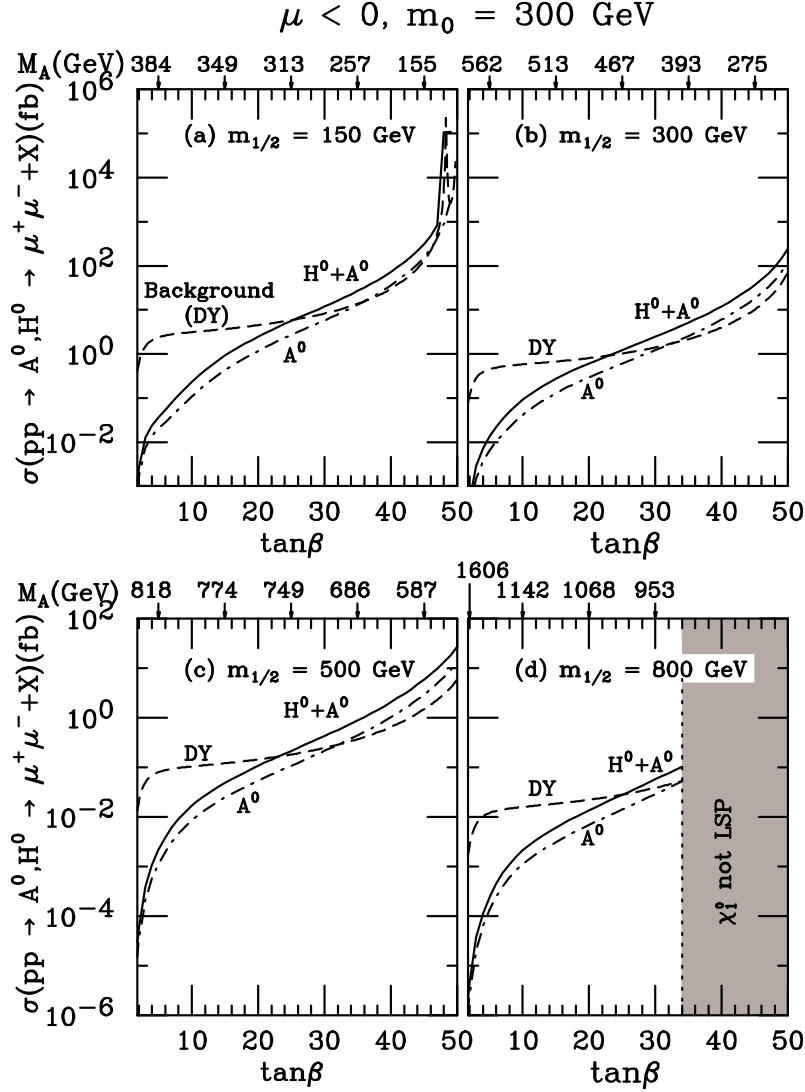}

\caption[]{
The total cross section of $pp \to A^0,H^0 \to \mu\bar{\mu} +X$ in fb 
at $\sqrt{s} = 14$ TeV, as a function of $\tan\beta$, 
for $\mu < 0$, $m_0 = 300$ GeV, and 
(a) $m_{1/2} = 150$ GeV, (b) $m_{1/2} = 300$ GeV, 
(c) $m_{1/2} = 500$ GeV, and, (d) $m_{1/2} = 800$ GeV, 
Also shown is the invariant mass distribution ($d\sigma/dM_{\mu\bar{\mu}}$)
of the Drell-Yan background (dashed) in fb/GeV for $M_{\mu\bar{\mu}} = m_A$.
The region with dark shading denotes the part of the parameter space 
that does not have the lightest neutralino ($\chi^0_1$) as the LSP.
\label{fig:tanb}
}\end{figure}

\begin{figure}
\centering\leavevmode
\epsfxsize=6in\epsffile{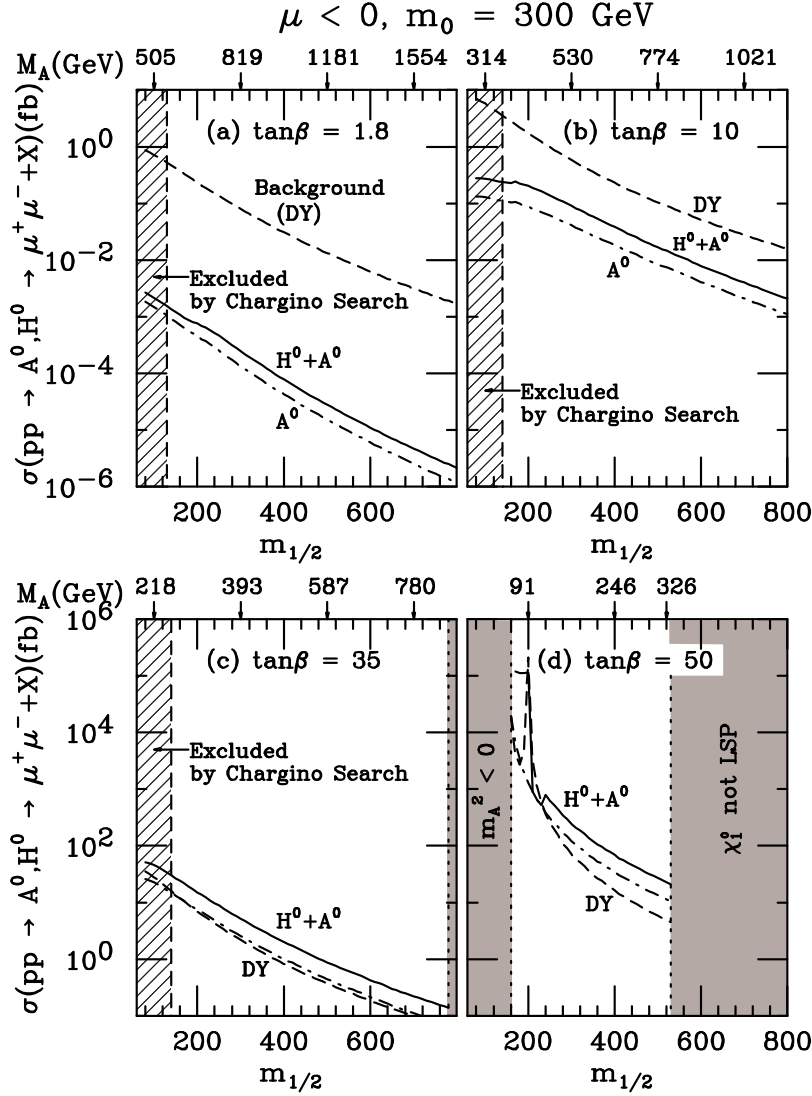}

\caption[]{
The total cross section of $pp \to A^0,H^0 \to \mu\bar{\mu} +X$ in fb 
at $\sqrt{s} = 14$ TeV, as a function of $m_{1/2}$, 
for $\mu < 0$, $m_0 = 300$ GeV, 
(a) $\tan\beta = 1.8$, 
(b) $\tan\beta = 10$, 
(c) $\tan\beta = 35$, and 
(d) $\tan\beta = 50$.  
Also shown is $d\sigma/dM_{\mu\bar{\mu}}$ 
of the Drell-Yan background (dashed) in (fb/GeV) for $M_{\mu\bar{\mu}} = m_A$.
The regions with dark shading are the parts of the parameter space 
excluded by theoretical requirements. 
The region excluded by the $m_{\chi^+_1} > 85$ GeV limit 
from the chargino search \cite{ALEPH} at LEP 2 is indicated.
\label{fig:mhf}
}\end{figure}

\begin{figure}
\centering\leavevmode
\epsfxsize=6in\epsffile{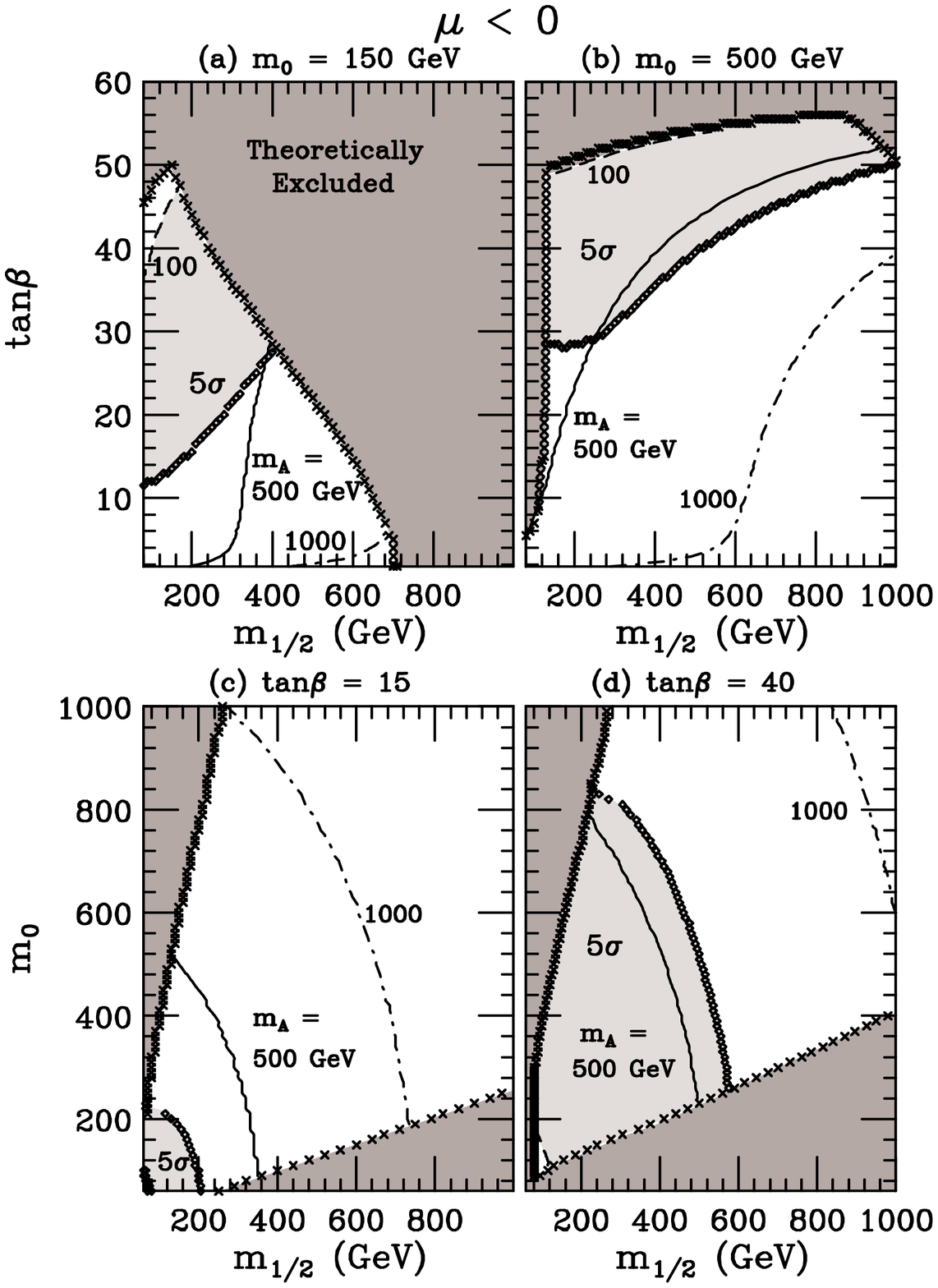}

\caption[]{
The $5\sigma$ contours at the LHC 
for an integrated luminosity ($L$) of 300 fb$^{-1}$ in 
(a) the $m_{1/2}$ versus $\tan\beta$ plane with $m_0 = 150$ GeV,
(b) the $m_{1/2}$ versus $\tan\beta$ plane with $m_0 = 500$ GeV,
(c) the $m_{1/2}$ versus $m_0$ plane with $\tan\beta = 15$, and 
(d) the $m_{1/2}$ versus $m_0$ plane with $\tan\beta = 40$. 
Also shown are the mass contours for $m_A = 100$ GeV (dashed), 
500 GeV (solid) and 1000 GeV (dot-dashed).
The discovery region is the part of the parameter space with light shading.
The regions with dark shading are the parts of the parameter space 
excluded by theoretical requirements. 
The region excluded by the $m_{\chi^+_1} > 85$ GeV limit 
from the chargino search \cite{ALEPH} at LEP 2 is indicated.
\label{fig:contour}
}\end{figure}
%


\begin{references}
\bibitem{MSSM}
H.P.~Nilles, Phys.~Rep. {\bf 110} (1984) 1; 
H.~Haber and G.~Kane, Phys.~Rep. {\bf 117} (1985) 75.
\bibitem{RGE} 
K.~Inoue, A.~Kakuto, H.~Komatsu and H.~Takeshita, 
Prog. Theor. Phys. {\bf 68} (1982) 927 and {\bf 71} (1984) 413.
\bibitem{Unification} 
P.~Langacker and M.~Luo, Phys. Rev. {\bf D44} (1991) 817;
J.~Ellis, S.~Kelley and D.~Nanopoulos, Phys. Lett. {\bf B260} (1991) 131; 
U.~Amaldi, W.~de Boer and H.~F\"urstenau, Phys. Lett. {\bf B260} (1991) 447; 
%
R.~Barbieri, 
talk given at the 18th International Symposium on Lepton-Photon Interactions, 
Hamburg, Germany, July 1997, hep-ph/9711232. 
\bibitem{BBO}
V.~Barger, M.S.~Berger, P.~Ohmann, 
Phys. Rev. {\bf D47} (1993) 1093; {\bf D49} (1994) 4908; 
V.~Barger, M.S.~Berger, P.~Ohmann and R.J.N.~Phillips, 
Phys. Lett. {\bf B314} (1993) 351.
\bibitem{IRFP}
B.~Pendleton and G.G.~Ross, Phys. Lett. {\bf B98} (1981) 291; 
C.T.~Hill, Phys. Rev. {\bf D24} (1981) 691; 
C.D.~Froggatt, R.G.~Moorhouse and I.G.~Knowles, 
Phys. Lett. {\bf B298} (1993) 356;   
%
J.~Bagger, S.~Dimopoulos and E.~Masso, 
Phys. Rev. Lett. {\bf 55} (1985) 920; 
%
H.~Arason, {\it et al.}, Phys. Rev. Lett. {\bf 67} (1991) 2933; 
and Phys. Rev. {\bf D46} (1992) 3945;
%
P.~Langacker, N.~Polonsky, Phys. Rev. {\bf D50} (1994) 2199; 
%
W.A.~Bardeen, M.~Carena, S. Pokorski and C.E.M.~Wagner, 
Phys. Lett. {\bf B320} (1994) 110; 
M.~Carena, M.~Olechowsk, S.~Pokorski and C.E.M.~Wagner 
Nucl. Phys. {\bf B419} (1994) 213;
%
B.~Schremp, Phys. Lett. {\bf B344} (1995) 193;
B.~Schrempp and M.~Wimmer, DESY-96-109 (1996), hep-ph/9606386. 
\bibitem{Guide} 
J.~Gunion, H.~Haber, G.~Kane and S.~Dawson, 
{\it The Higgs Hunter's Guide} (Addison-Wesley, Redwood City, CA, 1990).
\bibitem{HGG}
H.~Baer, M.~Bisset, C.~Kao and X.~Tata, Phys. Rev. {\bf D46} (1992) 1067.
\bibitem{Neutral}
V.~Barger, M.~Berger, A.~Stange and R.~Phillips, 
Phys. Rev. {\bf D45} (1992) 4128; 
J.~Gunion, R.~Bork, H.~Haber and A.~Seiden,
Phys. Rev. {\bf D46}, 2040 (1992); 
J.~Gunion, H.~Haber and C.~Kao, Phys. Rev. {\bf D46}, 2907 (1992); 
J.F.~Gunion and L.~Orr, Phys. Rev. {\bf D46} (1992) 2052. 
\bibitem{KZ}
Z.~Kunszt and F.~Zwirner, Nucl. Phys. {\bf B385} (1992) 3.
\bibitem{Z2Z2}
H.~Baer, M.~Bisset, D.~Dicus, C.~Kao and X.~Tata,
Phys. Rev. {\bf D47} (1993) 1062;
H.~Baer, M.~Bisset, C.~Kao and X.~Tata, Phys. Rev. {\bf D50} (1994) 316.
\bibitem{CMS}
CMS Technical Proposal, CERN/LHCC 94-38 (1994).
\bibitem{ATLAS}
Atlas Technical Proposal, CERN/LHCC 94-43 (1994).
\bibitem{ATLAS2}
E.~Richter-Was, D.~Froidevaux, F.~Gianotti, L.~Poggioli, D.~Cavalli, 
and S.~Resconi, CERN report CERN-TH-96-111, (1996).
\bibitem{AZh} 
S.~Abdullin, H.~Baer, C.~Kao, N.~Stepanov and X.~Tata, 
Phys. Rev. {\bf D54} (1996) 6728; 
H.~Baer, C.~Kao and X.~Tata, Phys. Lett. {\bf B303} (1993) 284. 
\bibitem{Dandi}
T.~Garavaglia, W.~Kwong and D.-D.~Wu, Phys. Rev. {\bf D48}, 1899 (1993).
\bibitem{DGV}
J.~Dai, J.F.~Gunion and R.~Vega, Phys. Lett. {\bf B315} (1993) 355;
Phys. Lett. {\bf B345} (1995) 29; {\bf B387} (1996) 801. 
\bibitem{hbb}
E.~Richter-Was and D.~Froidevaux, CERN report CERN-TH-97-210, (1997), 
hep-ph/9708455. 
\bibitem{Nikita}
C.~Kao and N.~Stepanov, Phys. Rev. D {\bf 52} (1995) 5025.
\bibitem{SUGRA} 
A.H.~Chamseddine, R.~Arnowitt and P.~Nath, 
Phys. Rev. Lett. {\bf 49} (1982) 970; 
L.~Iba\~nez and G.~Ross, Phys. Lett. {\bf B110} (1982) 215; 
L.~Iba\~nez, Phys. Lett. {\bf B118} (1982) 73; 
J.~Ellis, D.~Nanopoulos and K.~Tamvakis, Phys. Lett. {\bf B121} (1983) 123; 
L.~Alvarez-Gaum\'e, J.~Polchinski and M.~Wise, 
Nucl. Phys. {\bf B121} (1983) 495.
\bibitem{Non-universal} 
V.~Berezinskii et al., Astropart. Phys. {\bf 5} (1996) 1; 
P.~Nath and R.~Arnowitt, Northeastern Report No. NUB-TH-3151-97, (1997), 
hep-ph/9701301.
\bibitem{SUGRA2} 
J.~Ellis and F.~Zwirner, Nucl. Phys. {\bf B338} (1990) 317;
G.~Ross and R.G.~Roberts, Nucl. Phys. {\bf B377} (1992) 571; 
R.~Arnowitt and P.~Nath, Phys. Rev. Lett. {\bf 69} (1992) 725; 
M.~Drees and M.M.~Nojiri, Nucl. Phys. {\bf B369} (1993) 54; 
S.~Kelley {\it et. al.}, Nucl. Phys. {\bf B398} (1993) 3;
M.~Olechowski and S.~Pokorski, Nucl. Phys. {\bf B404} (1993) 590;
G.~Kane, C.~Kolda, L.~Roszkowski and J.~Wells, 
Phys. Rev. {\bf D49} (1994) 6173; 
D.J.~Casta\~no, E.~Piard and P.~Ramond, Phys. Rev. {\bf D49} (1994) 4882; 
W.~de~Boer, R.~Ehret and D.~Kazakov, Z. Phys. {\bf 67} (1995) 647;
%
H.~Baer, M.~Drees, C.~Kao, M.~Nojiri and X.~Tata, 
Phys. Rev. D {\bf 50} (1994) 2148; 
H.~Baer, C.-H.~Chen, R.~Munroe, F.~Paige and X.~Tata, 
Phys. Rev. D {\bf 51} (1995) 1046.
\bibitem{Baer} 
H.~Baer, C.-H.~Chen, M.~Drees, F.~Paige and X.~Tata, 
Phys. Rev. Lett. {\bf 79} (1997) 986.
\bibitem{Madison} 
V.~Barger and C.~Kao, University of Wisconsin report, MADPH-97-992, (1997), 
hep-ph/9704403. 
\bibitem{CLEO} 
M.S.~Alam et al., (CLEO Collaboration), Phys. Rev. Lett. {\bf 74} (1995) 2885.
\bibitem{LEP} 
P.G.~Colrain and M.I.~Williams, talk presented at 
the International Europhysics Conference on High Energy Physics, 
Jerusalem, Israel, August 1997. 
\bibitem{bsg} 
P.~Nath and R.~Arnowitt, Phys. Lett. {\bf B336} (1994) 395;
Phys. Rev. Lett. {\bf 74} (1995) 4592;
Phys. Rev. {\bf D54} (1996) 2374;
F.~Borzumati, M.~Drees and M.~Nojiri, Phys. Rev. {\bf D51} (1995) 341;
H.~Baer and M.~Brhlik, Phys. Rev. {\bf D55} (1997) 3201.
\bibitem{Higgs} 
H.~Haber and R.~Hempfling, Phys. Rev. Lett. {\bf 66} (1991) 1815;
J. Ellis, G. Ridolfi and F. Zwirner, Phys. Lett. {\bf B257} (1991) 83;
T. Okada, H. Yamaguchi and T. Tanagida, 
Prog. Theor. Phys. Lett. {\bf 85} (1991) 1; 
We use the calculations of 
M.~Bisset, Ph.D. thesis, University of Hawaii (1994).
\bibitem{Two-Loop} 
M.~Carena, J.R.~Espinosa, M.~Quiros, and C.E.M.~Wagner, 
Phys. Lett. {\bf B355} (1995) 209; 
M.~Carena, M.~Quiros, C.E.M.~Wagner, Nucl. Phys. {\bf B461} (1996) 407; 
H.~Haber, R.~Hempfling and A.~Hoang, CERN-TH/95-216 (1996), hep-ph/9609331.
\bibitem{CTEQ}
H.L.~Lai et al., Phys. Rev. {\bf D51} (1995) 4763.
\bibitem{Duane}
D.~Dicus and S.~Willenbrock, Phys. Rev. {\bf D39} (1989) 751.
\bibitem{QCD1} 
%
S.~Dawson, Nucl. Phys. {\bf B359} (1991) 283; 
%
A.~Djouadi, M.~Spira and P.M.~Zerwas, 
Phys. Lett. {\bf B264} (1991) 440;
D.~Graudenz, M.~Spira and P.M.~Zerwas, 
Phys. Rev. Lett. {\bf 70} (1993) 1372; 
M.~Spira, A.~Djouadi, D.~Graudenz and P.M.~Zerwas, 
Nucl. Phys. {\bf B453} (1995) 17. 
\bibitem{QCD2} 
S.~Dawson, A.~Djouadi and M.~Spira, 
Phys. Rev. Lett. {\bf 77} (1996) 16.
\bibitem{Collider}
V.~Barger and R.~Phillips, {\it Collider Physics, updated edition},  
(Addison-Wesley Publishing Company, Redwood City, CA, 1997).
\bibitem{Manuel} 
E.~Braaten, J.P.~Leveille, Phys. Rev. {\bf D22} (1980) 715; 
M.~Drees and K.~Hikasa, Phys. Lett. {\bf B240} (1990) 455; 
(E)-$ibid.$ {\bf B262} (1991) 497.
\bibitem{ALEPH}
ALEPH collaboration, talk presented at CERN by G.~Cowan, February, 1997.
\bibitem{Brown} N.~Brown, Z. Phys. {\bf C49} (1991) 657.
%
\end{references}
\end{document}